\documentclass[proof]{WileyASNA-v1}

\articletype{Article Type}%

\received{30 March 2022}
\revised{.. ...  2022}
\accepted{7 June  2022}

\begin{document}

\title{Optical spectroscopy of the Be/black hole binary MWC~656  - 
          interaction of a black hole with a circumstellar disc}

\author[1]{R. K. Zamanov*}

\author[1]{K. A. Stoyanov}

\author[2]{D. Marchev}

\author[1]{N. A. Tomov}

\author[3]{U. Wolter}

\author[4,5]{M. F. Bode}

\author[1]{Y. M. Nikolov}

\author[1]{S. Y. Stefanov}

\author[1]{A. Kurtenkov}

\author[1]{G. Y. Latev}

\authormark{Zamanov \textsc{et al}}

\address[1]{\orgdiv{Institute of Astronomy and National Astronomical Observatory}, \orgname{Bulgarian Academy of Sciences}, \orgaddress{\state{72 Tsarigradsko Shose, 1784 Sofia}, \country{Bulgaria}}}

\address[2]{\orgdiv{Department of Physics and Astronomy}, \orgname{Shumen University "Episkop Konstantin Preslavski"}, \orgaddress{\state{115 Universitetska Str., 9700 Shumen}, \country{Bulgaria}}}

\address[3]{\orgdiv{Hamburger Sternwarte}, \orgname{Universit\"at Hamburg}, \orgaddress{\state{Gojenbergsweg 112, 21029 Hamburg}, \country{Germany}}}

\address[4]{\orgdiv{Astrophysics Research Institute}, \orgname{Liverpool John Moores University}, 
\orgaddress{\state{IC2, 149 Brownlow Hill, Liverpool, L3 5RF}, \country{UK}}}

\address[5]{\orgdiv{Office of the Vice Chancellor}, \orgname{Botswana International University of Science and Technology}, \orgaddress{\state{Palapye}, \country{Botswana}}}

\corres{* \email{rkz@astro.bas.bg}}


\abstract{We present new spectroscopic observations of the Be/black hole binary MWC~656
  obtained during the period 2015 - 2021. We measure the equivalent width of 
  H$\alpha$ (EW$_\alpha$), H$\beta$ ($EW_\beta$), 
  and the distance between the peaks of  
  H$\alpha$ ($\Delta V_\alpha$), 
  H$\beta$ ($\Delta V_\beta$),  and FeII ($\Delta V_{FeII}$) lines. 
  Combining new and old data, we find that: 
  
  -- the density of the circumstellar disc of MWC~656 
     ($\Delta V_\alpha$ versus $EW_\alpha$ diagram) is similar to the Be stars. 
  For the Be stars we find the relation  
  $\Delta V_\beta = 0.999 \Delta V_\alpha + 62.4$~km~s$^{-1}$, 
  and the position of MWC~656 corresponds to the average behaviour of the Be stars.
  This means that the presence of the black hole does not change 
  the overall structure of the circumstellar disc. 
   
  -- the periodogram analysis indicates modulation of $EW_\alpha$ with a period
  $60.4 \pm 0.4$~days, which is identical to the binary orbital period.
  The maxima of $EW_\alpha$ and $EW_\beta$ are around  periastron (phase zero). 
    
  -- around orbital phase zero, 
  $\Delta V_\beta$ and $\Delta V_{FeII}$ decrease by about $30$~km~s$^{-1}$.  
  This suggests that we observe an increase of the circumstellar disc size 
  induced by the periastron passage of  the black hole and
  that the entire circumstellar disc 
  pulsates with the orbital period with 
  relative amplitude of 10-20\%.  
  
  The observations also indicate,  that the reason for the black hole in MWC~656 to be 
  in deep quiescence is a very low efficiency  of accretion ($\sim 2 \times 10^{-6}$).
}


\fundingInfo{Bulgarian National Science Fund -- project K$\Pi$-06-H28/2 08.12.2018  "Binary stars with compact object"}

\maketitle


\section{Introduction}
The Be/X-ray binaries consist of a rapidly rotating Be star and a
compact object accreting material from the circumstellar decretion disc. 
The primary has a  mass $> 8$~M$_\odot$  and the compact object can be
a neutron star or a black hole. 
The orbital periods are in the range 10 -- 400 d \citep{2011Ap&SS.332....1R}. 
The common understanding is that the neutron stars have a canonical mass of about 1.4~M$_\odot$ \citep{2000ARNPS..50..481H}. 
Recent studies support the idea that the neutron stars in high-mass X-ray binaries display
a relatively large range in mass -- from the theoretical lower mass 
limit of 1~M$_\odot$ up to over 2~M$_\odot$ \citep{2005ApJ...634.1242N,2019PhRvD.100b3015S}. 
The masses of the black holes in the high-mass X-ray binaries are in the range 
5 - 14~M$_\odot$ \citep{2020ApJ...898..143S}.  
The  Be/X-ray binaries are a product of the evolution of a binary containing 
two moderately massive stars, 
which undergoes mass transfer from the originally more massive star towards its companion \citep{1991A&A...241..419P,2007ASPC..367..477N}. 
The angular momentum transfer that accompanies the mass transfer 
onto the initially less massive star
is the most likely reason for the rapid rotation of the Be star \citep{2013ApJ...765....2P}. 
The eccentricities of the Be/X-ray binaries with orbital periods less than 150 days
are from 0 to 0.5 \citep{2018MNRAS.477.4810B,2019MNRAS.488..387B} 
and are caused by the natal kick to the compact object
and mass loss  during the
supernova explosion that formed it \citep[e.g.][]{1995MNRAS.274..461B,2009MNRAS.397.1563M}. 


The mass donors (primaries) of  the Be/X-ray binaries are emission-line Be stars. 
The Be stars are fast-rotating B-type stars with luminosity class III-V stars
which, at some point in their lives, have shown spectral lines
in emission \citep{2003PASP..115.1153P}. 
The material expelled from the equatorial belt of a rapidly rotating Be star forms an
outwardly diffusing gaseous, dust-free Keplerian disc \citep{2013A&ARv..21...69R}. 
In the optical band, the  most significant observational characteristic
of Be stars is the variable emission lines. 
These emission lines trace the presence of the surrounding decretion disc, 
and may appear and disappear together with the disc during the star's life.
Moving along the orbit, the compact object passes close to this disc, and sometimes 
may go through it. 
The gravitational force and tidal effect cause perturbations in the disc structure. 
This circumstellar disc feeds the accretion disc
around the compact object and/or interacts with its relativistic wind
and its magnetosphere.

 \begin{figure*}[ht!]     
  \vspace{15.0cm}   
  \includegraphics{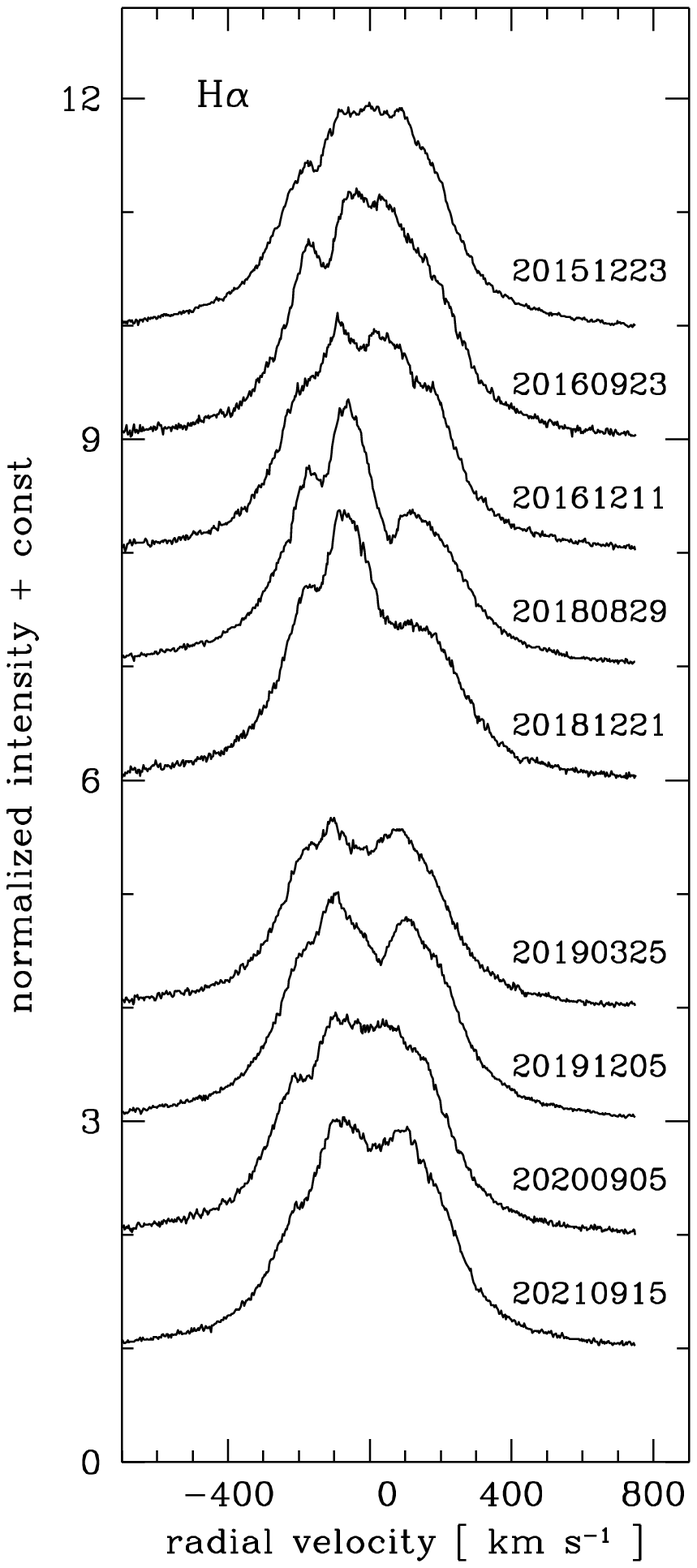}	  
  \includegraphics{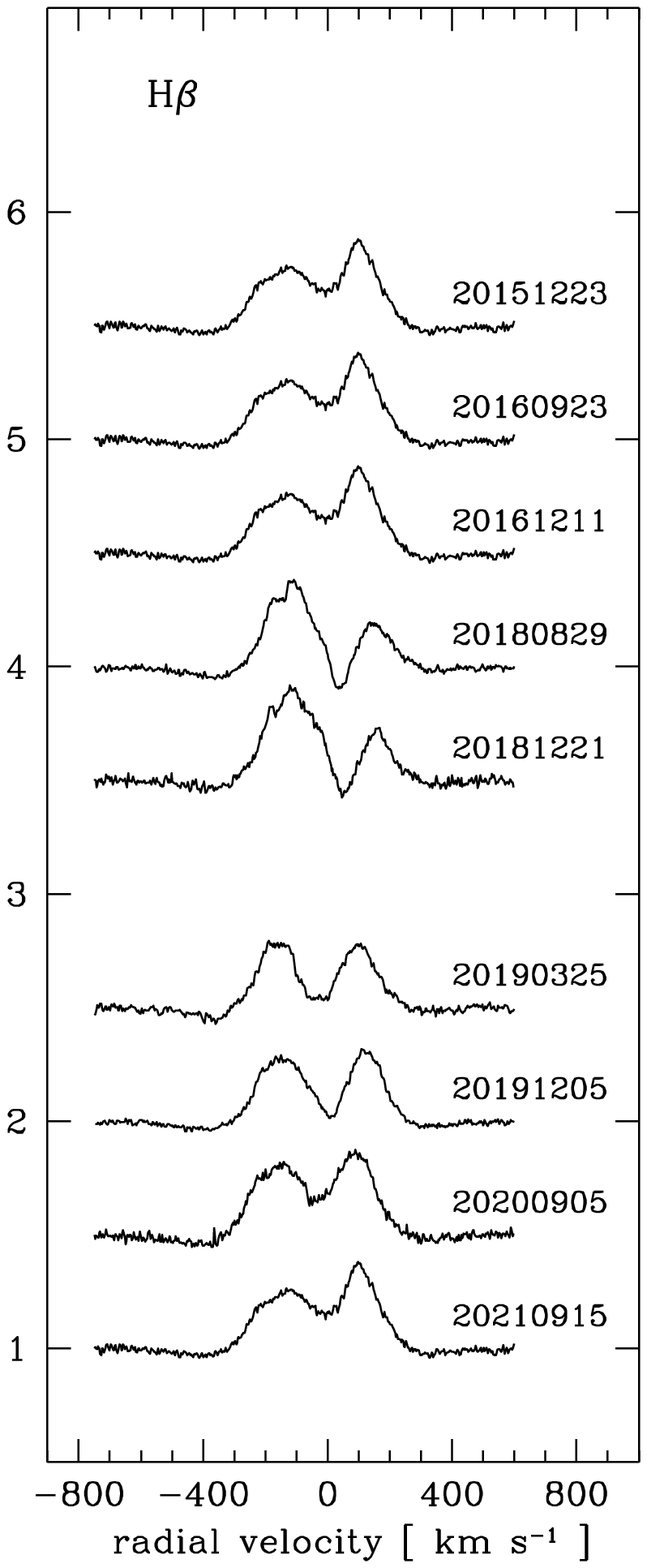}	  
  \includegraphics{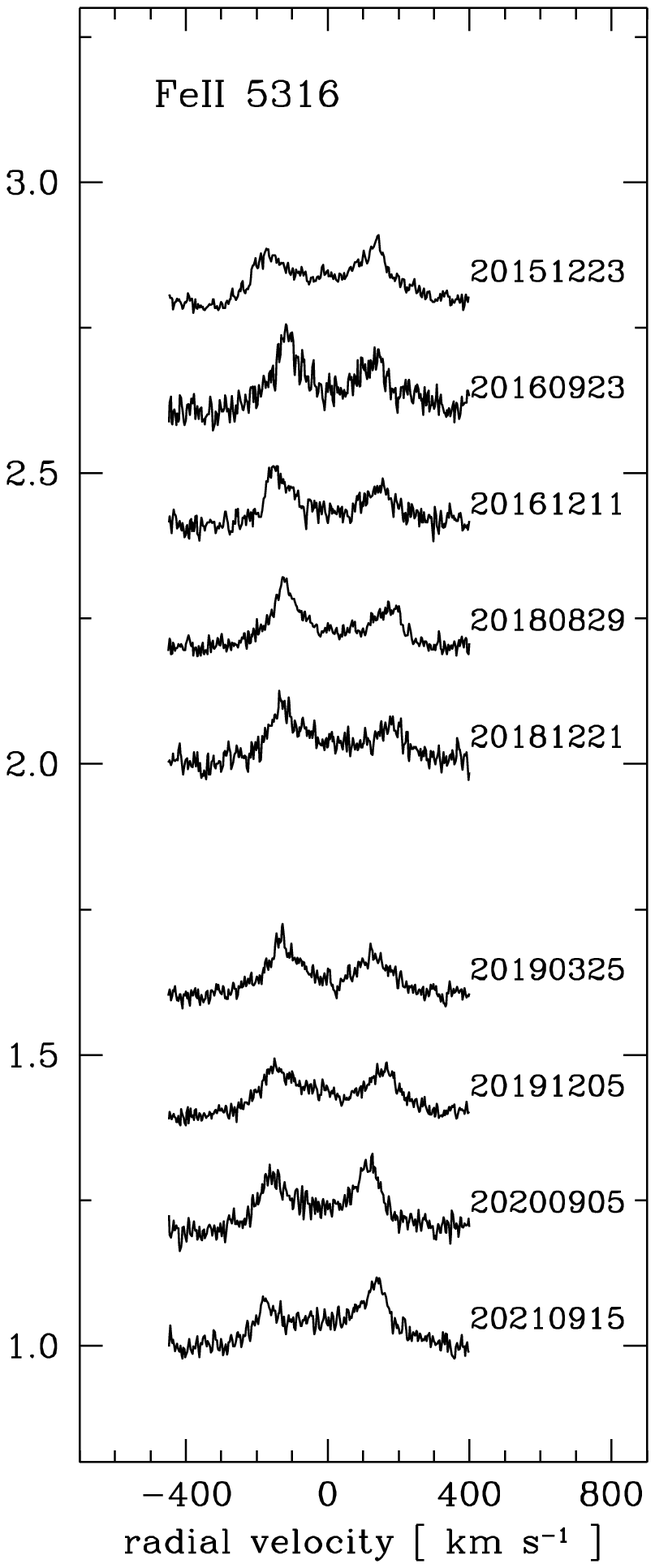}	  
  \caption[]{Variable profiles of the emission lines 
             H$\alpha$, H$\beta$ and FeII~5316 of MWC~656.
	     The date of observations is in the format YYYYMMDD. }
\label{f1.pro}      
\end{figure*}	     

 \begin{figure}[ht!]    
  \vspace{8.7cm} 
  \includegraphics{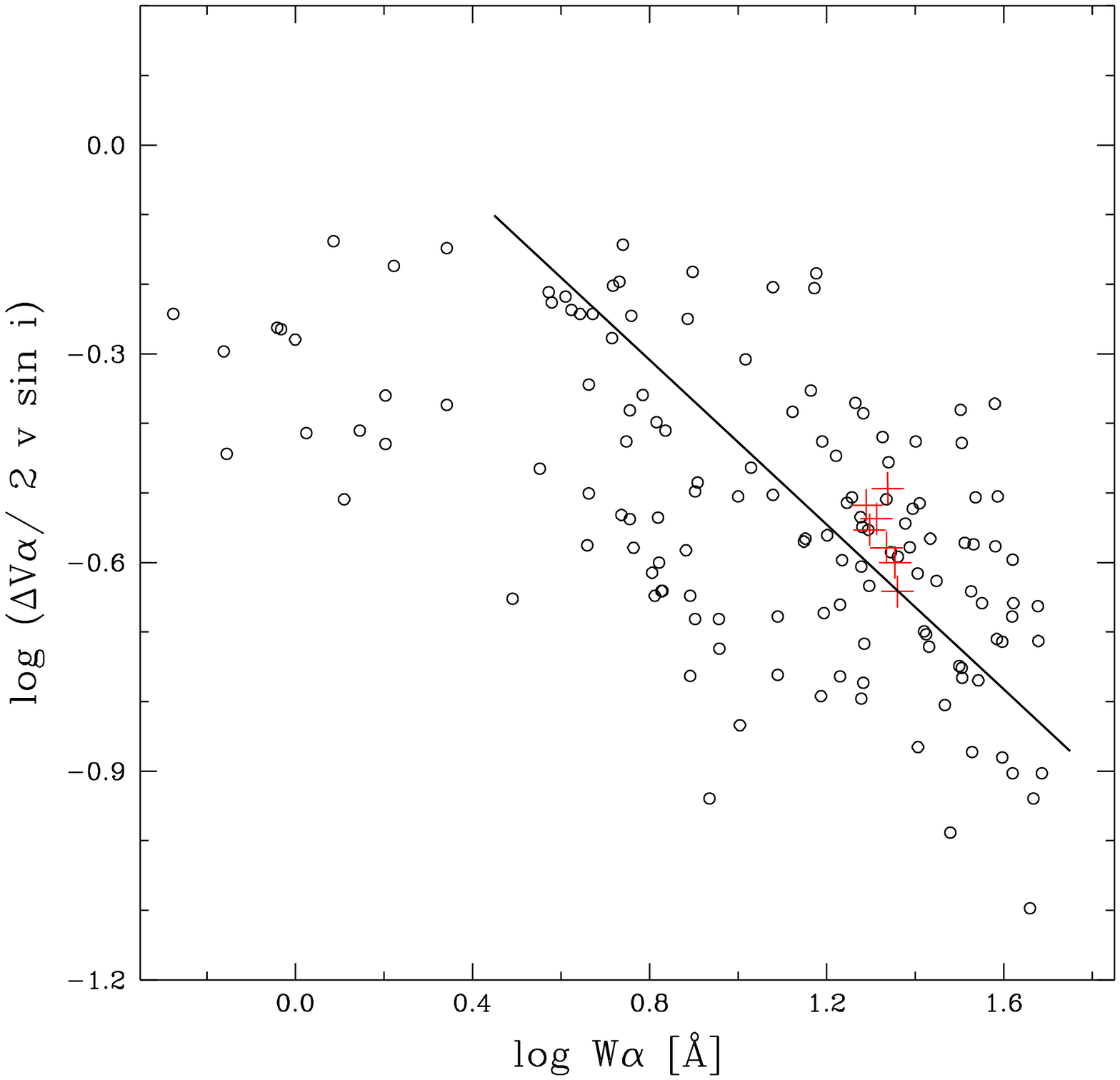}	  
  \caption[]{ $\Delta V_\alpha$ versus $EW_\alpha$. The black circles are the Be stars (see text)
  and the red plusses are our observations of MWC 656. The solid line is the average behaviour 
  of Be stars.}
  \label{f.Wa.dV}
  \vspace{8.9cm} 
  \includegraphics{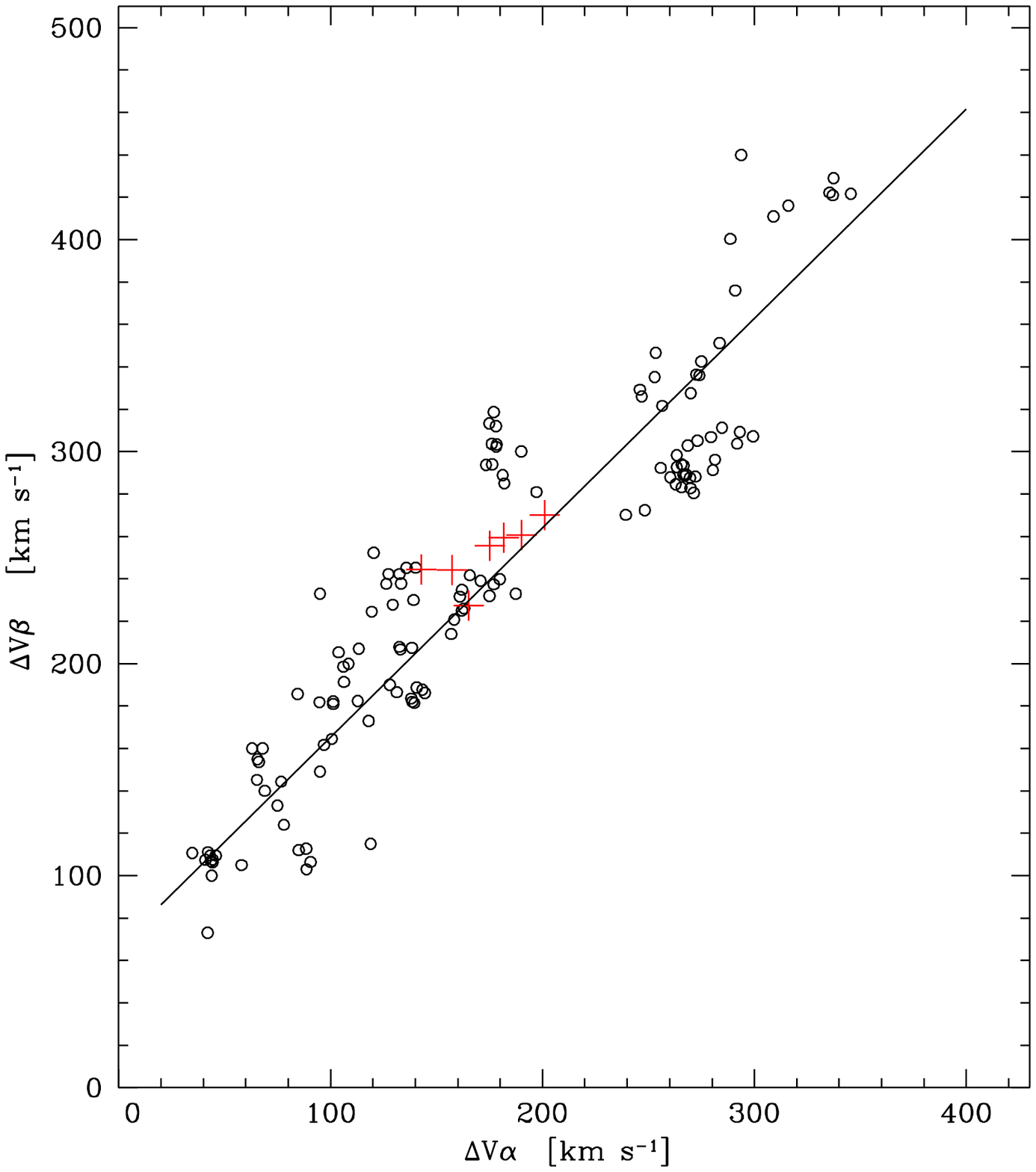} 	  
  \caption[]{ $\Delta V_\beta$ versus  $\Delta V_\alpha$ for Be stars (black circles).  
   The red crosses are our data for MWC~656. 
   The solid line is the linear fit 
   $\Delta V_\beta = 0.987 \; \Delta V_\alpha + 66.9$ km~s$^{-1}$.}
   \label{f.ab}  
\end{figure}
\begin{figure} 
  \vspace{9.2cm} 
  \includegraphics{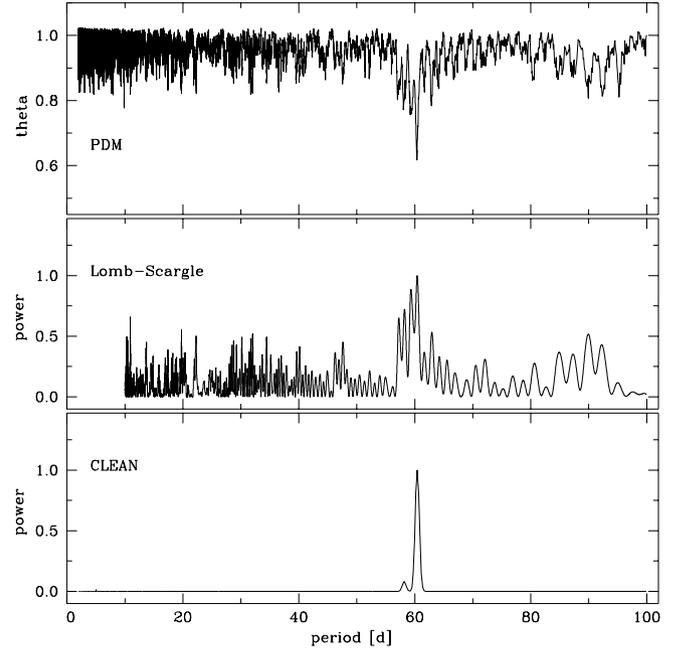}	  
  \caption[]{Periodogram analysis of $EW_\alpha$ with three independent methods
  (PDM, Lomb-Scargle and CLEAN). A periodic signal is clearly detected at the orbital period
  of the binary with PDM and CLEAN algorithms. }
  \label{f.pdm}
\end{figure}	    

\begin{figure} 
  \vspace{9.8cm} 
  \includegraphics{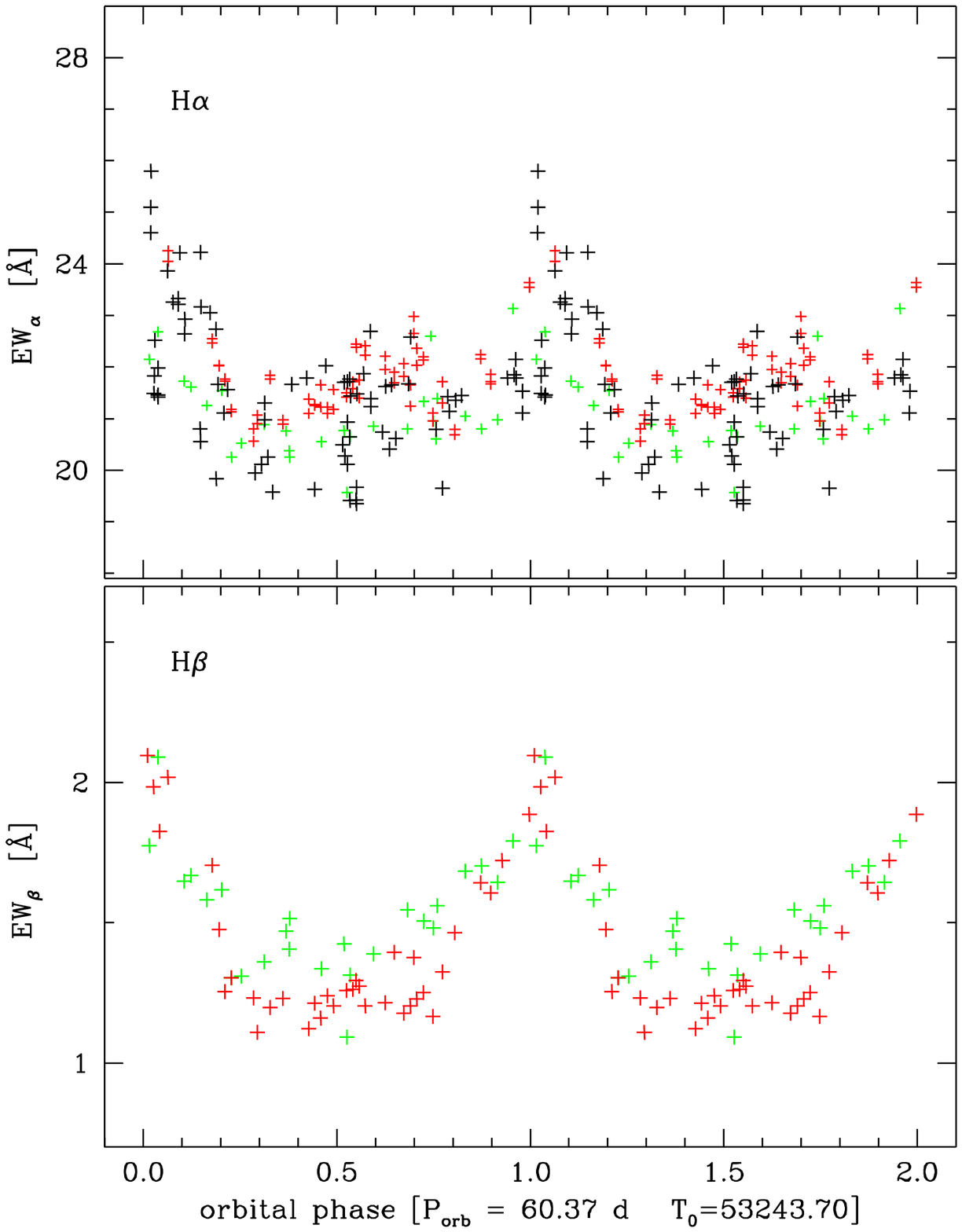}	  
  \caption[]{Orbital modulation of the equivalent widths of H$\alpha$ and H$\beta$ 
  emissions. 
  In Fig.~\ref{EWo} and Fig.~\ref{dVo} the folded light curves are shown twice for clarity. 
  }
  \label{EWo}  
  \vspace{9.8cm} 
  \includegraphics{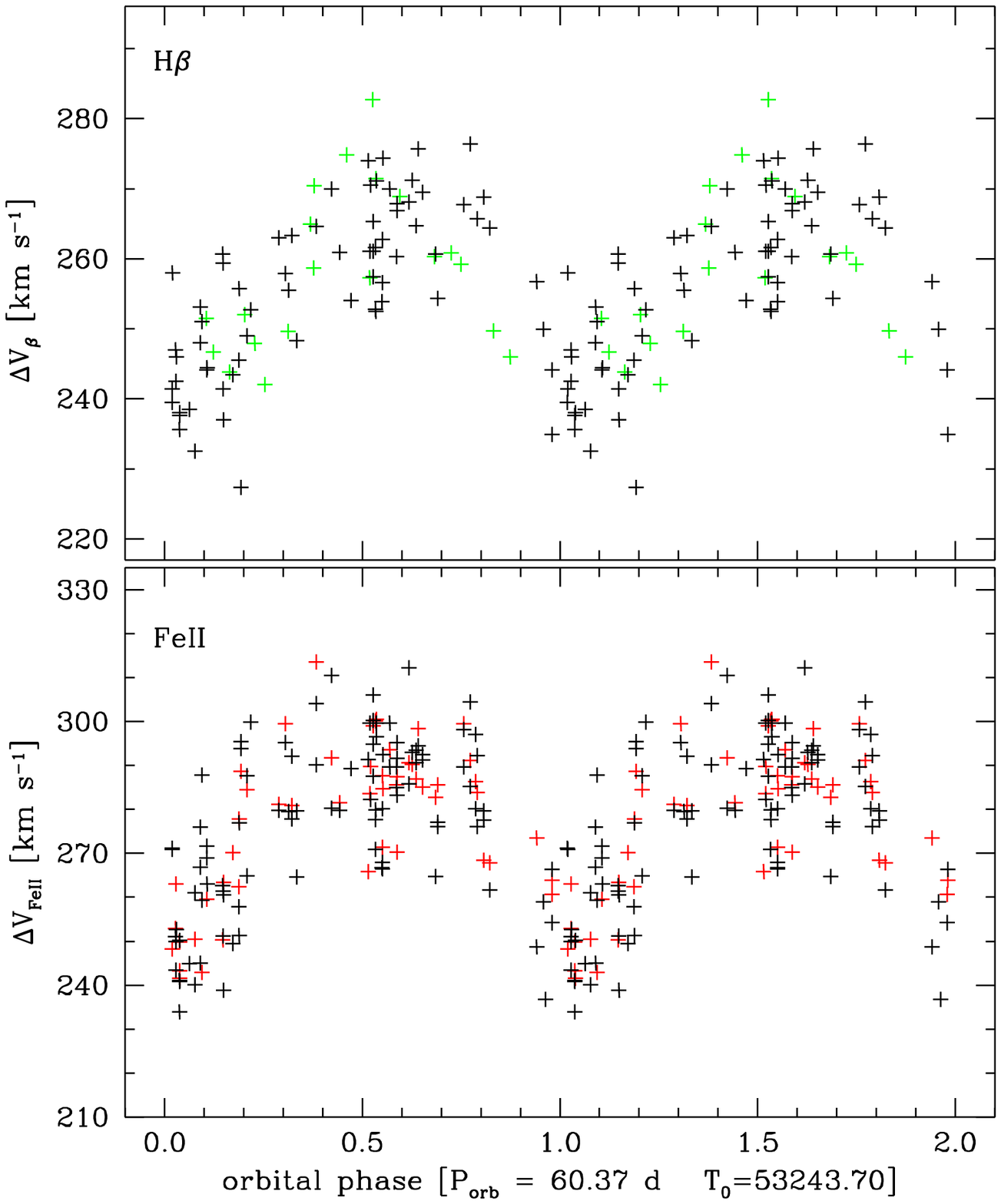}	  
  \caption[]{Orbital modulation of  $\Delta V_\beta$ and $\Delta V_{FeII}$.   
  Clear orbital modulation is visible in Fig.~\ref{EWo} and Fig.~\ref{dVo}, 
  indicating that the disc 
  is larger at  periastron (phase 0.0).
  }
  \label{dVo}
\end{figure}	    

MWC~656 (HD 215227) is the first binary system discovered to contain a black hole orbiting a Be star \citep{2014Natur.505..378C}. 
It is associated with the transient $\gamma$-ray source  AGL~J2241+4454 \citep{2010ATel.2761....1L}. 
MWC~656 is not included in 
the list of the confirmed $\gamma$-ray binaries because it was
only occasionally detected by the AGILE observatory at GeV energies 
and not yet detected in the TeV domain \citep{2015A&A...576A..36A}. 
The donor star in the system is a B1.5-2~III star
with a mass in the range 10 -- 16~M$_\odot$ \citep{2014Natur.505..378C}. 
The orbital period of the system is 60.37 $\pm$ 0.04~d, 
detected by optical photometry \citep{2010ApJ...723L..93W} 
and confirmed by radial velocity measurements \citep{2012MNRAS.421.1103C}. 
The spectral observations reveal the presence of a HeII~$\lambda$4686~\AA\ emission line produced by 
the accretion disc around the compact object. 
The analysis of the radial velocity of this line provides a secondary mass 
 3.8 -- 6.9~M$_\odot$, which proves that the compact object is a stellar mass  black hole \citep{2014Natur.505..378C}. 

Here we present optical spectroscopy of MWC~656  
from 2011-2021 and discuss the emission lines formed in the circumstellar disc 
and the orbital influence of the black hole on the Be star disc.

\section{Observations}
\label{s.obs}

We have obtained 73 optical spectra of MWC~656 on 50 nights secured with the 
ESpeRo Echelle spectrograph \citep{2017BlgAJ..26...67B}.   
on the 2.0 m RCC telescope of the Rozhen  
National Astronomical Observatory, Bulgaria. 6 of them are used in our previous studies. 
Some emission line profiles  of  H$\alpha$,  H$\beta$ and FeII~5316
of MWC~656 are presented in Fig.~\ref{f1.pro}. The spectra are normalized 
to the local continuum and a constant is added to each spectrum.
In this figure are plotted only 9 spectra. 
27 spectra were secured  with the HEROS spectrograph on the 1.2~m  TIGRE telescope \citep{2014AN....335..787S} 
in the astronomical observatory La Luz in Mexico during the period April 2019 - January 2020. 
In addition, we also use 65 red and 36 blue spectra [analyzed in \citet{2012MNRAS.421.1103C}] 
from the archive of the 2.0 m Liverpool Telescope \citep{2004SPIE.5489..679S}. 
These spectra were obtained using  
the Fibre-fed RObotic Dual-beam Optical Spectrograph [FRODOSpec; \citet{2004AN....325..215M}] 
during the period April-July 2011.

From the Rozhen spectra we measure the following parameters: 
the equivalent width of the  H$\alpha$ line (EW$_\alpha$),
the distance between the peaks of the following emission lines
H$\alpha$ ($\Delta V_\alpha$),
H$\beta$  ($\Delta V_\beta$),
$FeII~5316$ ($\Delta V_{5316}$),
$FeII~5197$ ($\Delta V_{5197}$),
and
$FeII~6435$ ($\Delta V_{6435}$).
To measure the position, we applied Gaussian fitting at the top of the peak. 
From the TIGRE spectra are measured EW$_\alpha$, EW$_\beta$,  and $\Delta V_\beta$,
and from the Liverpool Telescope spectra are measured EW$_\alpha$ and EW$_\beta$.
Because in the wings of H$\beta$, traces of the broad photospheric absorption
are visible, 
in all the cases EW$_\beta$ is the equivalent width of that part of the 
emission line which is above the level of the local continuum. 
The measurements are given in a few tables at the end of the paper -- 
Table~\ref{t.EWa} gives EW$_\alpha$,
Table~\ref{t.dVa} gives $\Delta V_\alpha$,
Table~\ref{t.EWb} gives EW$_\beta$,
Table~\ref{t.dVb} gives $\Delta V_\beta$, 
and Table~\ref{t.Fe} gives $\Delta V_{Fe}$.
The typical errors are $\pm 5$~\% for EW$_\alpha$,
$\pm 7$~\% for EW$_\beta$, $\pm 3$~km~s$^{-1}$ for $\Delta V_\beta$, 
and $\pm 5$~km~s$^{-1}$ for $\Delta V_{Fe}$.

\section{Results}
In only a few cases in our spectra does  the H$\alpha$ emission line show two peaks. 
H$\beta$ and Fe~II lines are double peaked on all the spectra. 
This is not surprising, as H$\alpha$ is optically thick and its shape is affected by radiative transfer effects (e.g. \citep[e.g.][]{1994A&A...289..458H,1994Ap&SS.216...87H} 
and by the presence of the black hole (see Sect.~\ref{orb.mod}). 

\subsection{Disc structure}

In Fig.~\ref{f.Wa.dV} we plot the distance between the peaks of H$\alpha$ normalized with the stellar 
rotation  versus $EW_\alpha$. These two parameters correlate in the Be stars, representing
the fact that $EW_\alpha$ increases as the disc grows. 
The black open circles are data for Be stars taken from 
\citet{1983A&AS...53..319A}, 
\citet{1986A&A...166..185H}, 
\citet{1988A&A...189..147H}, 
\citet{1992A&AS...95..437D}, 
\citet{1992ApJS...81..335S}  
and \citet{2013A&A...550A..79C}. 
The red plus signs are our measurements of MWC~656 when the H$\alpha$ emission is double peaked.

In Fig.~\ref{f.Wa.dV}, we see that MWC~656 lies within the Be stars population. The points lie
close to and slightly above the average line. 
For the projected rotational velocity of the primary component of MWC~656 
we used $v\,\sin{i}=313$~km~s$^{-1}$ \citep{2021AN....342..531Z}. 
The use of $v\,\sin{i}=330$~km~s$^{-1}$ \citep{2014Natur.505..378C}, 
will move the points slightly down and they will be even closer to the average line.
The vertical scatter of the points on this diagram is due to the density in the disc
[see Sect. 5.2 in \citet{1988A&A...189..147H}], 
which implies that the
density in the circumstellar disc in MWC~656 
is similar to the average of that in Be star discs.

In Fig.~\ref{f.ab} we plot  $\Delta V_\beta$ versus $\Delta V_\alpha$.
For this figure we used the data from \citet{1988A&A...189..147H}, 
our values for Be/$\gamma$-ray binaries \citep{2016A&A...593A..97Z} 
and our measurements on the spectra of \citet{2013A&A...550A..79C}. 
For the Be stars, \citet{1988A&A...189..147H} 
find that the peak separations of H$\beta$ and H$\alpha$ emission lines
follow  approximately the relation  $\Delta V_\beta \approx 1.8 \Delta V_\alpha$.
Using more data, we find that this relation is not valid above 
$\Delta V_\alpha \approx 200$~km~s$^{-1}$. 
Using a linear fit in the form $y=a+bx$,  we find the relation: 
\begin{equation}
\Delta V_\beta =  0.998 (\pm 0.008) \; \Delta V_\alpha + 62.4 (\pm 1.0) \; {\rm km \; s}^{-1},  
\end{equation}
where $\Delta V_\beta$ and $\Delta V_\alpha$ are in km~s$^{-1}$.
This relation is valid up to 360~km~s$^{-1}$. 
As expected there is a very strong correlation between  
these  two quantities, 
with correlation coefficient 0.91 and significance $10^{-10}$. 
The position of MWC~656 is very close to the average  behaviour of other Be stars.

The behaviour of MWC~656 in Fig.~\ref{f.Wa.dV} and Fig~\ref{f.ab}, 
indicates that the circumstellar disc is not strongly perturbed 
by the presence of the orbiting black hole 
and 
its overall structure is similar to the discs in the classical Be stars. 

\subsection{Disc size}
\label{Rd1}

For emission line profiles coming from a Keplerian disc,
the peak separation ($\Delta V$) can be regarded as a measure of 
the outer radius ($R_{disc}$) of the emitting disc \citep[e.g.][]{1972ApJ...171..549H}: 
 \begin{equation}
   \Delta V = 2 \; \sin{i} \; \sqrt{GM_1/R_{disc} \:}, 
  \label{H1}
  \end{equation}
where $G$ is the gravitational constant,  
$M_1$ is the mass of the Be star and 
$\sin{i}$ is the inclination of the disc to the line of sight. 
The projected rotational velocity ($v\,\sin{i}$) of a Be star 
can be expressed as
 \begin{equation}
   v \sin{i} = (1-\epsilon) \sin{i} \sqrt{G M_1/R_1 \:}, 
  \label{H2}
  \end{equation}
where $R_1$ is the radius of the Be star and 
$\epsilon$ is a dimensionless parameter 
for which we adopt $\epsilon = 0.1 \pm 0.1$.
The term $(1-\epsilon)$ represents the fact that 
the Be stars are rotating below the critical value \citep{2003PASP..115.1153P}. 
We obtain the following expression for $R_{d}$:
 \begin{equation}
   R_{disc} = R_1 \left( \frac{2 \; v \sin{i}}{(1-\epsilon) \; \Delta V}\right)^2, 
  \label{H3}
  \end{equation}
which relies on the assumptions that: 
(1) the disc is complanar with the equatorial plane of the Be   star,
(2) the Be star rotates below the critical rate, 
and
(3) the line profile shape is dominated by kinematics, 
and radiative transfer does not play a role.
Radius estimation through this method is a good approximation for symmetric profiles
with two peaks.

There is also a connection between  $R_{disc}$
and  EW$_\alpha$. We will use it in the form \citep[see also][]{2013A&A...559A..87Z}: 
 \begin{equation}
 R_{disc} = R_1 \; (1-\epsilon ) \; 0.467 \; EW_\alpha^{1.184}	
  \label{H4}
  \end{equation}
where $\epsilon$ is the same dimensionless parameter as in Eq.~\ref{H2}.  
A similar approach (with slightly different coefficients)
was applied by \citet{2006MNRAS.368..447C} 
and by \citet{2017MNRAS.464..572M}. 
Eq.~\ref{H4} gives us the possibility to estimate disc radius using measurements of EW$_\alpha$ 
and is applicable 
even in cases when the H$\alpha$ emission line does not display a double-peaked profile, but something different 
- single peak, triple peak profile, etc. 

Following Table~\ref{t.EWa}, we find that $EW_\alpha$ is  in the range $19.4  \le EW_\alpha  \le 25.8$~\AA\
with average value  $21.6 \pm 1.1$~\AA.
Using Eq.~\ref{H4},  $R_1=8.8$~R$_\odot$ \citep{2021AN....342..531Z}, 
and from the measurements of $EW_\alpha$, we estimate 
the size of the circumstellar disc in MWC~656 to be in the range 
 $123 \le  R_{disc} \le 174$~R$_\odot$, 
 with average value 
 $R_{disc} = 141 \pm 8$~R$_{\odot}$.

The distance between the peaks of H$\beta$ is in the range
$214 \le  \Delta V_\beta \le 282$~km~s$^{-1}$, with average value $255 \pm 14$~km~s$^{-1}$.
Using Eq.~\ref{H3} and our measurements of $\Delta V_\beta$, 
we find that 
 the  size of the disc visible in H$\beta$ emission
 is in the range $54 \le  R_{disc}(H\beta) \le 93$~R$_\odot$, 
 with average value 
 $R_{disc}(H\beta)=66 \pm 8$~R$_\odot$.

The distance between the peaks of the FeII~5316 line is in the range
$237 \le  \Delta V_{FeII5316} \le 305$~km~s$^{-1}$, with average value $274 \pm 19$~km~s$^{-1}$.
The distance between the peaks of the FeII~5197 line is in the range
$242 \le  \Delta V_{FeII5197} \le 314$~km~s$^{-1}$, with average value $277 \pm 18$~km~s$^{-1}$.
Finally, the distance between the peaks of the FeII~6433 line is in the range
$243 \le  \Delta V_{FeII5197} \le 312$~km~s$^{-1}$, with average value $285 \pm 18$~km~s$^{-1}$.
As can be expected, the three FeII lines have similar behaviour. 
Using Eq.~\ref{H3} and our measurements of $\Delta V_{FeII}$, 
we find that the size of the disc visible in FeII emission lines is
in the range $43 \le  R_{disc}(FeII) \le 76$~R$_\odot$, 
with average value $R_{disc}(FeII)= 56 \pm 8$~R$_\odot$.
The calculated 
disc sizes indicate that H$\alpha$ emission traces  the outer parts of the disc.
H$\beta$ and FeII emission lines are formed 
in a region much closer to the star than H$\alpha$,  
similar to the classical Be stars (see Sect.~5.3 in \citep{1988A&A...189..147H}). 

\subsection{Periodogram analysis}

The orbital motion of the compact object is expected to produce 
orbital modulation in the circumstellar disc. 
A summary of the observed orbital variations in three Be/$\gamma$-ray binaries 
is given in \citet{2021Univ....7..320M}. 
To search for signatures of the orbital modulation of the emission lines formed in the Be disc, 
we conducted a periodogram analysis of the measured emission line parameters (see Sect.~\ref{s.obs})
using the phase dispersion minimization, PDM \citep{1978ApJ...224..953S}, 
Lomb-Scargle \citep{1976Ap&SS..39..447L,1982ApJ...263..835S} 
and CLEAN \citep{1987AJ.....93..968R} 
algorithms. 
The analysis was performed  using the community-developed core Python package for
Astronomy \citep{2018AJ....156..123A}. 
The results of the periodogram analysis of $EW_\alpha$ are shown in Fig.~\ref{f.pdm}. 
The CLEAN algorithm shows a prominent peak at $60.43 \pm 0.44$~days 
and PDM statistics indicates a minimum at  $60.38 \pm 0.32$~days. 
The highest peak of Lomb-Scargle is at 60.45~days. 
The periodogram analysis of H$\beta$ parameters gives  similar values: 
$60.3 \pm 0.5$~days for EW$_\beta$  and  $60.5 \pm 0.7$~days  for $\Delta V_\beta$. 
Among the measured parameteres, the orbital period is more clearly  visible in EW$_\alpha$
probably because we have 165 measurements in total distributed over 10.5 years.   


\subsection{Orbital modulation}
\label{orb.mod}

Most of the Be/X-ray binaries show 
an asymmetric double-peaked H$\alpha$ line, 
usually caused by a distortion of the density distribution on the disc \citep{1997A&A...320..852H}. 
The variability of the emission lines is the main indicator 
for structural changes of the circumstellar disc. 
The variability of the H$\alpha$ emission reflects the changes in the outer parts of the disc, while  
the variability of  H$\beta$ and FeII emission lines - those in its inner parts.
An unperturbed circumstellar disc will produce a double-peaked emission line.
In only a few cases in our spectra does the H$\alpha$ line show two well defined peaks, in others
it appears as a single peak, profiles with 3 peaks, and flat top profiles 
(see the examples in Fig.~\ref{f1.pro}).
It means distortions of the density distribution in the disc \citep{1997A&A...320..852H}. 
H$\beta$ and FeII lines are always double-peaked in our spectra, which 
means  that the perturbations are stronger in the outer parts of the disc.
At distances less than 70~R$_\odot$ (which is approximately the size of the H$\beta$ disc) 
there are no large perturbations. 
This is in agreement with the supposition of \citet{2003PASP..115.1153P} 
that in the Be/X-ray binaries
the companion has little influence on the Be star and how its disc is produced, 
but alters only the
outer part of the disc. 
 

In Fig.~\ref{EWo} and Fig.~\ref{dVo} we plot a few parameters folded with the orbital 
period. We used  $P_{orb} = 60.37$~d
and $T_0 = 53243.70$. This ephemeris is from the analysis of the radial velocities \citep{2014Natur.505..378C} 
and the periastron is at phase~0.  
In each panel the folded light curve is shown twice for clarity. 
In Fig.~\ref{EWo} are plotted the orbital modulation of $EW_\alpha$ 
and $EW_\beta$.
The red pluses are the measurements from the Liverpool Telescope spectra, 
green are TIGRE, and black are Rozhen data. 
It is clear that the maxima of $EW_\alpha$ and $EW_\beta$
are  around the periastron when the EW$_\alpha$ is about 3~\AA\ larger. 
When the black hole is at apastron  EW$_\beta \approx 1.3$~\AA, while when 
at periastron it achieves $\approx 2$~\AA. 
When the black hole is at apastron  $EW_\alpha \approx 21$~\AA, while when 
at periastron  $\approx 24$~\AA.
Using Eq.~\ref{H4}, this corresponds to disc size increasing 
from 136~R$_\odot$ to 151~R$_\odot$. 
In other words the size of the H$\alpha$ disc pulsates 
with an amplitude of about  15~R$_\odot$ each orbital period. 
There is no noticeable time delay between the maxima of $EW_\alpha$ and $EW_\beta$.

In the upper panel of Fig.~\ref{dVo} is plotted 
the distance between the peaks of the H$\beta$ emission line.
The black symbols are Rozhen data, the green - TIGRE data.  
It is clear that  $\Delta V_\beta$ is larger at apastron and smaller 
during the periastron passage. Using Eq.~\ref{H3},
we calculate that changes of $\Delta V_\beta$ 
from 270~km~s$^{-1}$ to 240~km~s$^{-1}$ correspond 
to changes of the H$\beta$ disc from 47~R$_\odot$ to 60~R$_\odot$.

In the lower panel of Fig.~\ref{dVo} is plotted the variability of FeII lines. 
This panel contains measurements from the Rozhen spectra only. 
The black symbols indicate FeII~5316 and FeII~6433; the  red, the FeII~5197 line. 
It is evident that  $\Delta V_{FeII}$ is larger at apastron and smaller  
during the periastron passage. Using Eq.~\ref{H3},
we calculate that changes of $\Delta V_{FeII}$ 
from 300~km~s$^{-1}$ to 250~km~s$^{-1}$ correspond 
to changes of the FeII disc from 40~R$_\odot$ to 55~R$_\odot$.

The orbital modulation of the emission lines parameters 
plotted in Fig.~\ref{EWo} and Fig.~\ref{dVo}  indicates
that the  disc pulsates with orbital period. The pulsation is visible not only 
in the outer parts of the disc (H$\alpha$), but also in the inner parts
(H$\beta$ and FeII lines).  
The relative amplitude of the pulsations is 11\% for the H$\alpha$ emitting disc;
20\% for the H$\beta$ emitting disc, and 25\% for the FeII emitting disc.
It seems, that the relative amplitude of the pulsations 
decreases from the inner  toward the outer parts of the disc.
Usually the variability of quasi-Keplerian discs of the  Be  stars
has the form of a one-armed global disk oscillation (density and velocity wave)
as proposed by \citet{1997A&A...318..548O}. 
What we see in our observations of MWC~656 
is something different - pulsations 
induced by the orbital motion of a black hole.

\section{Discussion}

The interaction of the black hole in MWC~656 with the circumstellar disc 
and the accretion process are expected to be similar to that 
in the Be/X-ray binaries containing a neutron star. 
There are two points that we have to bear in mind about the influence of 
the compact object  
on the circumstellar disc:  
(i) a neutron star in a Be/X-ray binary can be in different regimes - 
accretor, propeller, ejector, georotator, etc.
\citep[e.g.][]{1987Ap&SS.132....1L}, 
while a black hole can only  be an accretor, 
(ii) the black hole is a few times more massive and 
its influence on the circumstellar disc will consequently 
be proportionately stronger.

For MWC~656, the orbital plane and the equatorial plane of the mass donor 
are probably complanar within  $\pm 4^0$ [see Sect. 5 in \citet{2021AN....342..531Z}]. 
Following this finding, using inclination $i=52^0 \pm 2^0$ \citep{2021AN....342..531Z}, 
and the values of \citet{2014Natur.505..378C} 
for the orbital eccentricity $e=0.1$, and for the semi-major axes
$a_1 \sin i = 38.0 \pm 6.3$~R$_\odot$  and 
$a_2 \sin i = 92.8 \pm 3.8$~R$_\odot$, we find 
the semi-major axis $a \approx 166$~R$_\odot$
and distance between components at periastron  $a(1-e) \approx 149.4$~R$_\odot$ 
and at apastron $a(1+e) \approx	182.6$~R$_\odot$. 

Using the formula by \citet{1983ApJ...268..368E} 
and mass ratio  $M_1/M_2 = 2.44$  \citep{2014Natur.505..378C}, 
we estimate the dimensionless Roche lobe size 
$r_L =  0.46$, 
and thus the Roche lobe size of the  primary would be 68~R$_\odot$ and 84~R$_\odot$, 
for periastron and apastron, respectively. 
The position of the inner Lagrangian point is 
at distance 88~R$_\odot$  and 108~R$_\odot$  for periastron and apastron, respectively.
This means that the circumstellar disc ($123 \le R_{disc} \le 174$~R$_\odot$, Sect.~\ref{Rd1}) 
is twice as large as the Roche lobe size and extends beyond the $L_1$ point. 
 
The interaction between the decretion disc and the neutron star  
in the Be/X-ray binaries results in
truncation of the disc \citep{2001A&A...377..161O,2002MNRAS.337..967O}. 
The observed orbital modulation of the emission lines of MWC~656 probably is due to 
periodic changes (oscillations) of the truncation radius as a result of the changes of the distance between the components. 
Other possibilities could be tidal waves induced by the black hole, 
or changes in the disc border, density and/or ionization induced by 
the gravitational force and high energy emission of the black hole.  


\subsection{Efficiency of accretion}

Typically the B and Be stars have mass loss rate values ranging from $10^{-11}$  to $3 \times 10^{-9}$
M$_\odot$~yr$^{-1}$  \citep{1981ApJ...251..139S}. 
More recently \citet{2017MNRAS.464.3071V} 
applied a viscous decretion disc model to 
the infrared disc continuum emission of 80 Be stars observed in different epochs
and estimated the disc mass decretion rates in the range between $10^{-12}$ 
and $10^{-9}$~M$_\odot$~yr$^{-1}$.
\citet{2018MNRAS.476.3555R} 
on the basis of a sample of 54 bright Be stars with clear disc events
find the typical mass loss rates associated with the disc events  of the order of $10^{-10}$~M$_\odot$~yr$^{-1}$.
It is worth noting that for 28~CMa, \citet{2012ApJ...744L..15C} 
find a higher mass injection rate $3.5 \pm 1.3 \times 10^{-8}\, M_\odot \,{yr}^{-1}$. 
Following the above and our results that the circumstellar disc of MWC~656 
is similar to the classical Be discs (Fig.~\ref{f.Wa.dV} and Fig.~\ref{f.ab}),
we can accept for the primary of MWC~656 a mass loss rate in the range 
$10^{-9} - 10^{-10}\,$~M$_\odot$~${yr}^{-1}$. 

The accretion radius of the black hole is:
\begin{equation}
R_a =  \frac{2 G M_{BH}}{V_{orb}^2 +c_s^2},
\end{equation}
where $V_{orb}$ is the orbital velocity of the black hole 
and $c_s$ is the speed of sound. We estimate R$_a \approx 250$~R$_\odot$.
Having this accretion radius, 
the black hole will be able to capture approximately 10-30\% of the mass loss of the primary. 
We expect a mass accretion rate onto the black hole to be  
$10^{-10} - 10^{-11}\,$~M$_\odot$~${yr}^{-1}$.
Another estimate can be obtained using 
the bright Be/X-ray binary X Per, where a 2.07~M$_\odot$  neutron star accretes 
from a Be star primary at a rate $1.5 \times 10^{-12} - 7 \times 10^{-12}\,$~M$_\odot$~yr$^{-1}$ 
\citep{2018PASJ...70...89Y,2019A&A...622A.173Z}. 
Bearing in mind that the orbital period is 4 times shorter and the black hole 
is 2.5 times more massive, we expect MWC~656 to exhibit at least 
a 5 times higher mass accretion rate,
$7.5 \times 10^{-12} - 3.5 \times 10^{-11}$. 
Both estimates imply  that the most likely  mass accretion rate 
onto the black hole of MWC~656 is 
$1-3 \times 10^{-11}\,$~M$_\odot$~${yr}^{-1}$. 

Multiwavelength observations reveal that MWC~656
is a faint X-ray \citep{2014ApJ...786L..11M} 
and radio \citep{2015A&A...580L...6D}. 
source.
\citet{2016ASPC..506..243A} 
place upper limits on the gamma-ray emission 
from this source and point out that it is not likely to be a true gamma-ray binary. 
\citet{2017ApJ...835L..33R} 
find that, 
MWC 656 is one of the faintest stellar-mass black holes ever detected in X-rays.
The luminosities and the obtained photon index are fully compatible with a black hole
in deep quiescence \citep{2013ApJ...773...59P}. 
\citet{2017ApJ...835L..33R} 
estimated X-ray luminosity $L_X \approx 3 \times 10^{30}$ erg s$^{-1}$ at a 2.6 kpc distance. 
Using the GAIA DR3 \citep{2021A&A...649A...1G} 
parallax $0.486$~mas,  
the model by \citet{2021AJ....161..147B} 
provides a distance 1985 pc. 
We recalculate   $L_X \approx 1.75 \times 10^{30}$ erg s$^{-1}$. 
The Eddington luminosity is $L_{Edd} = 1.26 \times 10^{38}$ M /M$_\odot$~erg~s$^{-1}$ for ionized Hydrogen.
We therefore estimate that the black hole in MWC~656 radiates at $2.8 \times 10^{-9}$ L$_{Edd}$. 

Following \citet{1973A&A....24..337S} 
and references therein, 
the total release of energy during the accretion is $L = \eta \dot M c^2$, 
where $\eta$ is the efficiency of gravitational energy release; 
in the case of Schwarzshild's metric $\eta \simeq 0.06$, 
in a Kerr black hole  $\eta$ can attain 40\%. 
Supposing that the black hole in MWC~656 accretes at a rate 
$1-3 \times 10^{-11}$~M$_\odot$~yr$^{-1}$, 
we find  $\eta \approx 2.10^{-6}$, which is  $\sim 3.10^4$
times lower than expected for a Schwarzshild's metric
and indicates that the likely reason the black hole to be in deep quiescence
is a very low efficiency of gravitational energy release.

%

\subsection{Similarity with the Be/neutron star binary PSR B1259-63}

PSR B1259-63/LS 2883 is one of the confirmed gamma-ray binary systems, consisting of
a  15-31~M$_\odot$ O9.5Ve star and a neutron star, visible as a 47.8 ms radio pulsar 
\citep{2011ApJ...732L..11N,2018MNRAS.479.4849M}. 
The radio observations of the pulsar allow the binary parameters to be well established -- 
1236.72 d ($\sim $3.4 yr) eccentric orbit, 
with an eccentricity of e = 0.87 \citep{2014MNRAS.437.3255S,2018MNRAS.479.4849M}. 
The spectral observations during three periastron passages \citep{2014MNRAS.439..432C,2015MNRAS.454.1358C,2021Univ....7..242C} 
and \citep{2016MNRAS.455.3674V} 
reveal an increase of $EW_\alpha$
at the time of the periastron of the neutron star.
There is also a hint of a decrease of the distance between 
the peaks of HeI at the time of the periastron of the neutron star
[see Fig.~1a in \citet{2016MNRAS.455.3674V} 
and Fig.~1d in \citet{2021Univ....7..242C}]. 
The behaviour of $EW_\alpha$ is very similar during the three periastron passages.  

Moreover, during the periastron passage, PSR B1259-63/LS 2883 gets 
brighter in the NIR bands \citep{2021PASJ...73..545K}. 
The observed increase of the $EW_\alpha$ and  
NIR variability are likely 
due to an expansion of the circumstellar disc caused by the tidal force of the compact object.

The behaviour of PSR B1259-63/LS 2883 
around the periastron resembles our results for MWC~656 -- an increase of 
$EW_\alpha$ and decrease of the $\Delta V$ (H$\beta$ and FeII) at the periastron 
(Fig.~\ref{EWo} and Fig.~\ref{dVo}).

\subsection{Comparison with LS~I~+61~303}

LS~I~+61~303 is a Be/gamma-ray binary with periodic outbursts and
orbital period 26.496~days [\citet{2002ApJ...575..427G} 
and references therein]. The primary is a B0Ve star. 
The secondary is most probably a neutron star \citep{2022NatAs.tmp...71W}, 
however some models consider the possibility of a black hole \citep{2017MNRAS.468.3689M,2015A&A...580L...6D}.
The orbital modulation of H$\alpha$ is visible in the ratio
of the equivalent widths of  the blue and red humps, and   
is an indication of the redistribution of the material in the disc - part of the time 
there is more material in the blue side and part of the time more material in the red side. 
The peak of EW$_\alpha$ is not at the periastron, 
it is in between periastron and apastron [see Fig.~1 in \citet{2013A&A...559A..87Z}]. 
This is different from the behaviour of PSR B1259-63 and MWC~656. 

It is worth noting, that the derived position of the periastron of LS~I~+61~303 is
from the orbital solutions of \citet{2005MNRAS.360.1105C} 
and \citet{2009ApJ...698..514A}, 
based on the radial velocities. 
Some recent models by \citet{2020A&A...643A.170K} 
and \citet{2022A&A...658A.153C} 
cast doubts on the position of the periastron. 
However the radial velocity measurements are model independent and more reliable.

The measurements of EW$_\alpha$ of MWC~656 cover 10 years from April 2011
to November 2021. During this decade no big variations of the EW$_\alpha$
are detected,  it varies in a relatively narrow range $19.4 \le EW_\alpha \le 25.8$~\AA. 
During 25 years of H$\alpha$ observations of LS~I~+61~303 also 
no large variations of the EW$_\alpha$ are observed \citep{2013A&A...559A..87Z}. 
In both stars the circumstellar disc is stable and no episode of disc loss is observed.
A difference is that in LS~I~+61~303 the H$\alpha$ emission is always double-peaked, while 
in MWC~656 it is double-peaked on only a few occasions, which means that
it is  perturbed. A possible reason can be that in MWC~656 the black hole 
is more massive than the neutron star in LS~I~+61~303. 

We note in passing, that it will be interesting 
to study the inner disc of LS~I~+61~303, which probably is seen in H$\beta$ and H$\gamma$ lines
and to observe PSR B1259-63  during one or two entire orbital periods. 


\section{Conclusions}
We analysed 165 spectra of the Be/black hole  binary MWC~656 
obtained with Rozhen, TIGRE and Liverpool telescopes
during the period April 2011 - October 2021. 
We studied the orbital modulation of the H$\alpha$,
H$\beta$, and FeII emission lines, which are formed in the Be circumstellar
disc. The orbital modulations of the emission lines suggest pulsations
in the circumstellar disc induced by the orbital motion of the black hole. 
The overall structure of the circumstellar 
disc is similar to that of the Be stars and 
the reason the black hole in MWC~656 appears to be
in deep quiescence is probably a very low efficiency of accretion.


\section*{Acknowledgments}
This work was supported by the  \fundingAgency{Bulgarian National Science Fund}   
\fundingNumber{project KP-06-H28/2 08.12.2018  "Binary stars with compact object"}.
TIGRE is a collaboration of the Hamburger Sternwarte, the Universities of Hamburg, 
Guanajuato and Liege.
Liverpool Telescope is operated on the island of La Palma by Liverpool John
Moores University in the Spanish Observatorio del Roque
de los Muchachos of the Instituto de Astrofisica de Canarias
with financial support from the UK STFC.
DM acknowledges support from the Research Fund of the Shumen University.



\subsection*{Conflict of interest}

The authors declare no potential conflict of interests.

\bibliography{zamanovref}

\jnlcitation{\cname{%
\author{R. K. Zamanov},
\author{K. A. Stoyanov},
\author{D. Marchev},
\author{N. A. Tomov},
\author{U. Wolter},
\author{M. F. Bode},
\author{Y. M. Nikolov},
\author{S. Y. Stefanov}
\author{A. Kurtenkov}, and
\author{G. Y. Latev}} (\cyear{2022}), 
\ctitle{Optical spectroscopy of the Be/black hole binary MWC~656  - 
          interaction of a black hole with a circumstellar disc}, 
\cjournal{Astronomische Nachrichten / Astronomical Notes}, \cvol{2022;00:1--6}.}

 \section*{Supporting information}
The following supporting information: Table~\ref{t.EWa} EW$_\alpha$, Table~\ref{t.dVa}  $\Delta V_\alpha$,
Table~\ref{t.EWb}  EW$_\beta$,
Table~\ref{t.dVb} $\Delta V_\beta$, 
and Table~\ref{t.Fe}  $\Delta V_{Fe}$), is available as part of the online article:

\begin{table*}
\centering
\caption{EW$_\alpha$ of MWC~656. In the Table are given HJD, EW$_\alpha$, and the telescope
Ro=Rozhen, TI=TIGRE, LT=Liverpool telescope. 
}  
\begin{tabular}{ccc | ccc | ccc | ccc}
\hline 
HJD-2450000  & EW$_\alpha$[\AA] &  & HJD   & EW$_\alpha$  &  & HJD    & EW$_\alpha$ &  &  HJD    & EW$_\alpha$   \\
             &                  &  &       &              &  &        &             &  &         &               \\
7209.45820   & 	22.58  &  Ro  &  8889.22715  &  20.49  &  Ro  & 8778.5527   &  20.80  &  TI & 5746.53928  &  21.65  &  LT  \\
7238.51411   & 	23.05  &  Ro  &  9071.39779  &  21.77  &  Ro  & 8787.6122   &  21.05  &  TI & 5746.54302  &  21.22  &  LT  \\
7239.43916   & 	22.73  &  Ro  &  9071.41932  &  20.65  &  Ro  & 8798.6807   &  22.15  &  TI & 5747.58250  &  21.10  &  LT  \\
7380.23985   &	21.70  &  Ro  &  9071.44710  &  19.41  &  Ro  & 8807.6478   &  21.25  &  TI & 5747.58623  &  21.22  &  LT  \\
7381.22172   &	21.44  &  Ro  &  9072.43325  &  19.42  &  Ro  & 8816.5657   &  20.88  &  TI & 5748.52945  &  21.56  &  LT  \\
7382.17444   &	21.48  &  Ro  &  9072.45478  &  19.67  &  Ro  & 8825.5471   &  20.55  &  TI & 5748.53318  &  21.18  &  LT  \\
7383.23478   &	21.87  &  Ro  &  9072.47701  &  19.35  &  Ro  & 8833.6131   &  20.85  &  TI & 5750.54930  &  21.50  &  LT  \\ 
7384.24999   &	22.69  &  Ro  &  9097.33973  &  22.15  &  Ro  & 8842.5554   &  22.60  &  TI & 5750.55303  &  21.42  &  LT  \\ 
7418.24782   &	23.16  &  Ro  &  9097.36264  &  21.79  &  Ro  & 8843.5732   &  21.39  &  TI & 5751.53900  &  21.58  &  LT  \\ 
7478.55794   &	24.22  &  Ro  &  9098.39114  &  21.11  &  Ro  & 8865.5772   &  21.61  &  TI & 5751.54273  &  21.71  &  LT  \\ 
7558.39879   &  22.02  &  Ro  &  9098.41336  &  21.53  &  Ro  & 5675.70236  &  20.80  &  LT & 5752.51401  &  21.40  &  LT  \\ 
7561.42121   &  20.28  &  Ro  &  9101.27107  &  21.49  &  Ro  & 5675.70609  &  20.56  &  LT & 5752.51773  &  21.75  &  LT  \\ 
7654.52680   &  23.86  &  Ro  &  9101.29259  &  21.83  &  Ro  & 5691.71744  &  22.45  &  LT & 5753.49472  &  22.23  &  LT  \\ 
7655.36222   &  23.26  &  Ro  &  9101.38912  &  22.52  &  Ro  & 5691.72118  &  22.38  &  LT & 5753.49844  &  22.41  &  LT  \\ 
7734.23018   &  21.66  &  Ro  &  9126.32609  &  19.63  &  Ro  & 5697.64370  &  21.69  &  LT & 5756.57664  &  21.95  &  LT  \\ 
8267.52316   &  21.56  &  Ro  &  9162.22581  &  21.42  &  Ro  & 5697.64743  &  21.90  &  LT & 5756.58037  &  22.21  &  LT  \\ 
8274.55830   &  19.58  &  Ro  &  9162.25984  &  21.98  &  Ro  & 5700.70738  &  22.65  &  LT & 5759.49896  &  21.82  &  LT  \\ 
8360.48052   &  20.79  &  Ro  &  9162.29109  &  21.46  &  Ro  & 5700.71112  &  22.98  &  LT & 5759.50269  &  22.07  &  LT  \\ 
8361.44027   &  19.65  &  Ro  &  9277.14227  &  21.79  &  Ro  & 5703.68169  &  21.11  &  LT & 5760.52770  &  21.65  &  LT  \\ 
8362.52294   &  21.14  &  Ro  &  9278.14016  &  21.85  &  Ro  & 5703.68542  &  20.95  &  LT & 5760.53142  &  21.24  &  LT  \\ 
8363.53339   &  21.36  &  Ro  &  9358.51178  &  19.95  &  Ro  & 5712.71041  &  21.86  &  LT & 5761.52250  &  22.03  &  LT  \\ 
8364.45841   &  21.45  &  Ro  &  9359.52225  &  20.11  &  Ro  & 5712.71414  &  21.67  &  LT & 5761.52623  &  22.36  &  LT  \\ 
8473.34350   &  21.62  &  Ro  &  9360.50008  &  20.25  &  Ro  & 5712.71414  &  21.72  &  LT & 5762.54107  &  22.14  &  LT  \\ 
8474.22399   &  21.65  &  Ro  &  9420.35524  &  20.97  &  Ro  & 5718.70789  &  23.55  &  LT & 5762.54480  &  22.20  &  LT  \\ 
8567.62876   &  19.83  &  Ro  &  9420.38997  &  21.30  &  Ro  & 5718.71162  &  23.64  &  LT & 5765.51294  &  21.30  &  LT  \\
8625.51303   &  20.80  &  Ro  &  9473.43785  &  21.66  &  Ro  & 5722.70951  &  24.25  &  LT & 5765.51667  &  21.71  &  LT  \\ 
8625.55539   &  20.55  &  Ro  &  9474.34828  &  21.11  &  Ro  & 5722.71323  &  24.05  &  LT & 5767.46415  &  20.69  &  LT  \\ 
8648.45537   &  21.71  &  Ro  &  9509.27856  &  21.43  &  Ro  & 5729.63596  &  22.47  &  LT & 5767.46788  &  20.79  &  LT  \\ 
8648.46301   &  20.93  &  Ro  &  8601.9518   &  20.60  &  TI  & 5729.63968  &  22.55  &  LT & 5771.49588  &  22.24  &  LT  \\ 
8648.47065   &  20.11  &  Ro  &  8613.9492   &  23.13  &  TI  & 5730.66177  &  22.02  &  LT & 5771.49969  &  22.16  &  LT  \\ 
8682.42632   &  23.22  &  Ro  &  8622.9631   &  21.73  &  TI  & 5730.66551  &  22.03  &  LT  \\ 
8682.46869   &  23.33  &  Ro  &  8631.9509   &  20.52  &  TI  & 5731.61273  &  21.73  &  LT  \\ 
8683.44166   &  22.64  &  Ro  &  8648.9298   &  20.65  &  TI  & 5731.61645  &  21.77  &  LT  \\ 
8683.48403   &  22.93  &  Ro  &  8661.8871   &  20.95  &  TI  & 5732.60107  &  21.18  &  LT  \\ 
8714.34399   &  20.74  &  Ro  &  8671.8757   &  20.98  &  TI  & 5732.60480  &  21.13  &  LT  \\ 
8715.41556   &  20.41  &  Ro  &  8690.7678   &  20.25  &  TI  & 5736.66567  &  20.90  &  LT  \\ 
8716.38644   &  20.61  &  Ro  &  8699.7475   &  20.38  &  TI  & 5736.66940  &  21.07  &  LT  \\ 
8718.34346   &  21.67  &  Ro  &  8699.8523   &  20.25  &  TI  & 5738.64170  &  21.77  &  LT  \\ 
8803.41370   &  24.21  &  Ro  &  8708.8044   &  19.57  &  TI  & 5738.64543  &  21.84  &  LT  \\ 
8823.24312   &  21.79  &  Ro  &  8720.7572   &  21.33  &  TI  & 5740.65714  &  20.89  &  LT  \\ 
8833.22997   &  21.39  &  Ro  &  8729.7546   &  20.80  &  TI  & 5740.66086  &  20.97  &  LT  \\ 
8833.23830   &  21.23  &  Ro  &  8739.6932   &  22.68  &  TI  & 5744.66896  &  21.38  &  LT  \\ 
8859.24002   &  24.60  &  Ro  &  8749.6375   &  21.54  &  TI  & 5744.67277  &  21.10  &  LT  \\ 
8859.27613   &  25.10  &  Ro  &  8759.5988   &  20.76  &  TI  & 5745.59023  &  21.25  &  LT  \\ 
8859.30390   &  25.80  &  Ro  &  8768.6880   &  20.77  &  TI  & 5745.59396  &  21.27  &  LT  \\ 
 \\			    
 \hline 	     		        		    		     
 \end{tabular}       		        			     
 \label{t.EWa}    		        	   
 \end{table*}	     				    

\begin{table}[t!]
\centering
\caption{$\Delta V_\alpha$.  } 
\begin{tabular}{ccccccc}
\hline 
HJD-2450000  & $\Delta V_\alpha$  &      &  \\
            & [km s$^{-1}$]      &      &  \\ 
            &                    &      &  \\ 
8567.62876  &   175.2            &  Ro  &  \\
8625.51314  &	190.2            &  Ro  &  \\  
8625.55570  &	181.8            &  Ro  &  \\  
8683.44186  &	157.3            &  Ro  &  \\  
8683.48414  &	142.9            &  Ro  &  \\  
8823.24315  &	201.1            &  Ro  &  \\  
9473.43827  &	165.2            &  Ro  &  \\  
\\			    	   
 \hline 	     		        		    		     
 \end{tabular}       		        			     
 \label{t.dVa}    		        	   
 \end{table}	     				    
\begin{table}
\centering
\caption{$EW_\beta$. } 
\begin{tabular}{ccc  | ccc r  }
\hline 
HJD-2450000  & $EW_\beta$ & & HJD-2450000  & $EW_\beta$ &  \\  
             & [ \AA]     & &   	  & [ \AA]     &  \\ 
	     &            & &   	  &            &  \\
5675.69897   & 1.232   & LT & 8613.94918  &    1.791 &  TI  &  \\
5691.71464   & 1.294   & LT & 8622.96308  &    1.648 &  TI  &  \\ 
5697.64122   & 1.394   & LT & 8631.95087  &    1.310 &  TI  &  \\ 
5700.70501   & 1.376   & LT & 8648.92984  &    1.314 &  TI  &  \\ 
5703.67952   & 1.165   & LT & 8661.88711  &    1.481 &  TI  &  \\ 
5712.70860   & 1.607   & LT & 8671.87572  &    1.644 &  TI  &  \\
5718.70660   & 1.887   & LT & 8690.76781  &    1.303 &  TI  &  \\ 
5722.70830   & 2.018   & LT & 8699.74746  &    1.406 &  TI  &  \\ 
5729.63535   & 1.705   & LT & 8699.85233  &    1.515 &  TI  &  \\ 
5730.66109   & 1.476   & LT & 8708.80439  &    1.092 &  TI  &  \\ 
5731.61209   & 1.254   & LT & 8720.75715  &    1.505 &  TI  &  \\ 
5732.60065   & 1.303   & LT & 8729.75464  &    1.703 &  TI  &  \\ 
5736.66551   & 1.109   & LT & 8739.69322  &    2.091 &  TI  &  \\ 
5738.64154   & 1.197   & LT & 8749.63753  &    1.618 &  TI  &  \\  
5740.65725   & 1.230   & LT & 8759.59880  &    1.471 &  TI  &  \\  
5744.66932   & 1.122   & LT & 8768.68801  &    1.424 &  TI  &  \\  
5745.59068   & 1.212   & LT & 8778.55269  &    1.546 &  TI  &  \\  
5746.53979   & 1.159   & LT & 8787.61219  &    1.684 &  TI  &  \\  
5747.58307   & 1.239   & LT & 8798.68068  &    1.774 &  TI  &  \\  
5748.53010   & 1.204   & LT & 8807.64779  &    1.582 &  TI  &  \\  
5750.55008   & 1.259   & LT & 8816.56567  &    1.361 &  TI  &  \\  
5751.53967   & 1.262   & LT & 8825.54706  &    1.335 &  TI  &  \\  
5752.51489   & 1.273   & LT & 8833.61308  &    1.388 &  TI  &  \\  
5753.49567   & 1.204   & LT & 8843.57318  &    1.560 &  TI  &  \\  
5756.57774   & 1.215   & LT & 8865.57720  &    1.669 &  TI  &  \\  
5759.50029   & 1.176   & LT & \\  		        
5760.52910   & 1.204   & LT & \\  		        
5761.52397   & 1.228   & LT & \\  
5762.54258   & 1.250   & LT & \\  
5765.51464   & 1.324   & LT & \\  
5767.46596   & 1.464   & LT & \\  
5771.49789   & 1.642   & LT & \\  
6076.70430   & 1.722   & LT & \\  
6081.70660   & 2.096   & LT & \\  
6082.68514   & 1.984   & LT & \\  
6083.60974   & 1.825   & LT & \\  
\\			 	    
 \hline 	     	 						      
 \end{tabular}       	 					      
 \label{t.EWb}    		        	   
 \end{table}	     				    

\begin{table*}
\centering
\caption{$\Delta V_\beta$. In the first part are given Rozhen, in the second TIGRE data. } 
\begin{tabular}{cc | cc | cc cc}
\hline 
HJD-2400000  & $\Delta V_\beta$ & HJD   &  $\Delta V_\beta$ &  HJD   &  $\Delta V_\beta$ \\
             & [km s$^{-1}$]    &       &                   & \\
                                                           		&	     \\
 Rozhen      &             &    Rozhen     &          &      TIGRE	&	     \\
 57209.45820 &     254.3   &   58859.24002 &    239.5 &     58622.96308 &     251.5  \\
 57238.51411 &     243.4   &   58859.27613 &    241.4 &     58631.95087 &     242.0  \\
 57239.43916 &     245.5   &   58859.30390 &    258.0 &     58648.92984 &     271.4  \\
 57380.23985 &     261.1   &   58889.22715 &    274.0 &     58661.88711 &     259.2  \\
 57381.22172 &     271.1   &   59071.39779 &    252.8 &     58671.87572 &     188.1  \\ 
 57382.17444 &     274.4   &   59071.41932 &    261.6 &     58690.76781 &     247.9  \\ 
 57383.23478 &     270.0   &   59071.44710 &    252.5 &     58699.74746 &     258.7  \\ 
 57384.24999 &     260.3   &   59072.43325 &    253.9 &     58699.85233 &     270.4  \\ 
 57418.24782 &     237.0   &   59072.45478 &    256.6 &     58708.80439 &     282.7  \\ 
 57478.55794 &     241.4   &   59072.47701 &    262.8 &     58720.75715 &     260.8  \\ 
 57558.39879 &     254.0   &   59098.39114 &    244.1 &     58729.75464 &     246.0  \\ 
 57561.42121 &     270.5   &   59098.41336 &    234.9 &     58739.69322 &     219.5  \\    
 57654.52680 &     238.5   &   59101.27107 &    247.0 &     58749.63753 &     252.0  \\ 
 57655.36222 &     232.5   &   59101.29259 &    242.5 &     58759.59880 &     264.9  \\ 
 57734.23018 &     264.6   &   59101.38912 &    246.0 &     58768.68801 &     257.3  \\
 58267.52316 &     252.7   &   59126.32609 &    260.9 &     58778.55269 &     260.3  \\ 
 58274.55830 &     248.3   &   59162.22581 &    235.6 &     58787.61219 &     249.7  \\ 
 58360.48052 &     267.7   &   59162.25984 &    237.6 &     58798.68068 &     264.7  \\ 
 58361.44027 &     276.4   &   59162.29109 &    238.0 &     58807.64779 &     243.8  \\ 
 58362.52294 &     265.7   &   59277.14227 &    256.7 &     58816.56567 &     249.6  \\ 
 58363.53339 &     268.8   &   59278.14016 &    249.9 &     58825.54706 &     274.8  \\ 
 58364.45841 &     264.4   &   59358.51178 &    263.0 &     58833.61308 &     268.9  \\ 
 58473.34350 &     271.2   &   59359.52225 &    257.9 &     58843.57318 &     241.9  \\ 
 58474.22399 &     275.7   &   59360.50008 &    263.3 &     58865.57720 &     246.7  \\ 
 58567.62876 &     255.7   &   59420.38997 &    255.5 &   \\
 58625.51303 &     260.7   &   59473.43785 &    227.3 &   \\ 
 58625.55539 &     259.4   &   59474.34828 &    241.5 &   \\ 
 58648.45537 &     257.4   &    \\ 
 58648.46301 &     265.3   &    \\ 
 58648.47065 &     261.1   &    \\ 
 58682.42632 &     248.0   &    \\ 
 58682.46869 &     253.1   &    \\ 
 58683.44166 &     244.1   &    \\ 
 58683.48403 &     244.4   &    \\ 
 58714.34399 &     268.1   &    \\ 
 58715.41556 &     264.7   &    \\ 
 58716.38644 &     269.5   &    \\ 
 58718.34346 &     260.7   &    \\ 
 58803.41370 &     251.0   &    \\ 
 58823.24312 &     270.0   &    \\ 
 58833.22997 &     266.9   &    \\ 
 58833.23830 &     267.9   &    \\  
 \\			    				       
 \hline 	     		        		    		     
 \end{tabular}       		        			     
 \label{t.dVb}    		        	   
 \end{table*}	     				    

\begin{table*}
\centering
\caption{$\Delta V_{Fe}$.} 
\begin{tabular}{cc | cc | cc | cc  cccccccc}
\hline 
 HJD-2400000 & $\Delta V_{Fe}$ & HJD        & $\Delta V_{Fe}$ & HJD      & $\Delta V_{Fe}$ & HJD      & $\Delta V_{Fe}$ \\
             & [km s$^{-1}$]   & 	    &                 & 	 &                 & 	      &                 \\
             &                 & 	    &	              & 	 &  	           & 	      &                 \\
 FeII 5316   &         &   FeII 5316  &         &    FeII 6433  &	  &  FeII~5197   &	   &  \\
 57209.45820 &   276.1 &  58889.22715 &   291.4 &   58715.41556 &   293.5 &  58833.23830 &   287.4 &  \\  
 57238.51411 &   249.4 &  59071.39779 &   270.9 &   58716.38644 &   292.5 &  58859.27613 &   248.1 &  \\ 
 57239.43916 &   257.9 &  59071.41932 &   277.7 &   58803.41370 &   287.8 &  58889.22715 &   265.9 &  \\ 
 57380.23985 &   299.6 &  59071.44710 &   279.9 &   58823.24312 &   310.5 &  59071.41932 &   300.1 &  \\ 
 57381.22172 &   299.6 &  59072.43325 &   266.7 &   58833.23830 &   295.2 &  59072.45478 &   271.4 &  \\ 
 57382.17444 &   292.5 &  59072.45478 &   266.5 &   59072.45478 &   280.1 &  59072.47701 &   287.6 &  \\ 
 57383.23478 &   289.6 &  59072.47701 &   268.0 &   59101.29259 &   249.9 &  59098.39114 &   260.7 &  \\ 
 57384.24999 &   289.7 &  59098.39114 &   254.2 &   59162.22581 &   241.1 &  59098.41336 &   263.8 &  \\ 
 57418.24782 &   238.9 &  59098.41336 &   266.4 &   59360.50008 &   292.1 &  59101.27107 &   252.9 &  \\ 
 57478.55794 &   251.2 &  59101.27107 &   251.0 &   59473.43785 &   293.9 &  59101.29259 &   263.0 &  \\ 
 57558.39879 &   289.3 &  59101.29259 &   243.4 &   59474.34828 &   264.9 &  59126.32609 &   281.5 &  \\ 
 57561.42121 &   282.2 &  59101.38912 &   252.6 &   59509.27856 &   280.1 &  59162.22581 &   249.8 &  \\ 
 57654.52680 &   244.9 &  59126.32609 &   279.8 & 		&	  &  59162.25984 &   243.2 &  \\ 
 57655.36222 &   240.1 &  59162.22581 &   233.9 &    FeII~5197  &	  &  59162.29109 &   241.5 &  \\ 
 57734.23018 &   290.1 &  59162.25984 &   240.8 &   57209.45820 &   285.6 &  59277.14227 &   273.5 &  \\ 
 58267.52316 &   299.9 &  59162.29109 &   250.2 &   57238.51411 &   270.1 &  59358.51178 &   281.0 &  \\ 
 58274.55830 &   264.6 &  59277.14227 &   248.7 &   57239.43916 &   262.4 &  59359.52225 &   299.5 &  \\ 
 58360.48052 &   289.7 &  59278.14016 &   259.0 &   57380.23985 &   283.6 &  59360.50008 &   280.8 &  \\ 
 58361.44027 &   285.2 &  59358.51178 &   279.8 &   57381.22172 &   300.5 &  59473.43785 &   288.7 &  \\ 
 58362.52294 &   292.3 &  59359.52225 &   295.2 &   57382.17444 &   284.6 &  59474.34828 &   284.5 &  \\  
 58363.53339 &   279.7 &  59360.50008 &   277.8 &   57383.23478 &   293.6 &  59509.27856 &   286.3 &  \\  
 58364.45841 &   261.7 &  59420.38997 &   279.5 &   57384.24999 &   285.6 &  \\ 
 58473.34350 &   293.0 &  59473.43785 &   295.5 &   57418.24782 &   263.3 &  \\ 
 58474.22399 &   294.5 &  59474.34828 &   287.7 &   57561.42121 &   289.8 &  \\ 
 58567.62876 &   251.3 &  59509.27856 &   297.0 &   57655.36222 &   250.3 &  \\ 
 58625.51303 &   262.6 & 	      & 	&   57734.23018 &   313.5 &  \\ 
 58625.55539 &   261.4 &  FeII 6433   & 	&   58360.48052 &   299.5 &  \\  
 58648.45537 &   287.6 &  57209.45820 &   277.1 &   58361.44027 &   291.1 &  \\  
 58648.46301 &   300.2 &  57381.22172 &   296.6 &   58362.52294 &   283.8 &  \\  
 58648.47065 &   294.8 &  57383.23478 &   299.7 &   58363.53339 &   268.4 &  \\  
 58682.42632 &   266.8 &  57384.24999 &   285.0 &   58364.45841 &   267.8 &  \\  
 58682.46869 &   245.0 &  57418.24782 &   260.5 &   58473.34350 &   290.2 &  \\  
 58683.44166 &   269.0 &  57655.36222 &   261.0 &   58474.22399 &   298.4 &  \\ 
 58683.48403 &   263.0 &  57734.23018 &   304.1 &   58567.62876 &   277.8 &  \\ 
 58714.34399 &   285.8 &  58274.55830 &   279.7 &   58625.51303 &   250.2 &  \\ 
 58715.41556 &   290.6 &  58360.48052 &   298.1 &   58648.47065 &   298.9 &  \\ 
 58716.38644 &   291.2 &  58361.44027 &   304.4 &   58683.44166 &   259.5 &  \\ 
 58718.34346 &   264.8 &  58362.52294 &   276.1 &   58714.34399 &   290.6 &  \\ 
 58803.41370 &   259.3 &  58363.53339 &   277.6 &   58715.41556 &   286.9 &  \\ 
 58823.24312 &   280.3 &  58567.62876 &   276.9 &   58716.38644 &   285.1 &  \\ 
 58833.22997 &   291.6 &  58648.47065 &   306.0 &   58718.34346 &   282.7 &  \\ 
 58833.23830 &   283.2 &  58682.42632 &   276.0 &   58803.41370 &   242.9 &  \\ 
 58859.24002 &   271.2 &  58683.44166 &   271.7 &   58823.24312 &   291.7 &  \\ 
 58859.27613 &   270.9 &  58714.34399 &   312.2 &   58833.22997 &   270.3 &  \\ 
 \\
 \hline 			      
 \end{tabular}  		     
 \label{t.Fe}		
 \end{table*}					    

\end{document}